\def \yskip{\penalty-50\vskip3pt plus 3pt minus 2pt}
\def \reference{\par \yskip \noindent \hangindent .4in \hangafter 1}
\def \abc#1#2#3#4 {\reference#1, {\sl#2}, {\bf#3}, #4}
\def \blank {\lower 5pt\hbox to 0.75in{\hrulefill}}

\def \cm{~\rm{cm}}
\def \s{~\rm{s}}
\def \km{~\rm{km}}

\def \AU{~\rm{AU}}
\def \yrs{~\rm{yrs}}
\def \yr{~\rm{yr}}

\def \G{~\rm{G}}
\def \erg{~\rm{erg}}

\def \lae{\mathrel{<\kern-1.0em\lower0.9ex\hbox{$\sim$}}}
\def \gae{\mathrel{>\kern-1.0em\lower0.9ex\hbox{$\sim$}}}

\documentstyle[11pt,aasms,tighten,flushrt]{article}
\begin{document}
%\normalsize
\small

\setcounter{page}{1}
%\noindent Presented at the 180 IAU Symposium: {\it Planetary Nebulae}, 
%August 1996.
%dust1.tex
\begin{center} \bf 
SOLAR-LIKE CYCLE IN ASYMPTOTIC GIANT BRANCH STARS
\end{center}
%\vspace*{2.0cm}

\begin{center}
Noam Soker\\
Department of Physics, University of Haifa at Oranim\\
%Mathematics-Physics\\
Oranim, Tivon 36006, ISRAEL \\
soker@physics.technion.ac.il 
\end{center}

%\clearpage 

\begin{center}
\bf ABSTRACT
\end{center}

 I propose that the mechanism behind the formation of 
concentric semi-periodic shells found in several planetary nebulae (PNs)
and proto-PNs, and around one asymptotic giant branch (AGB) star,
is a solar-like magnetic activity cycle in the progenitor AGB stars.
The time intervals between consecutive ejection events is
$\sim 200-1,000 \yrs$, which is assumed to be the cycle period
(the full magnetic cycle can be twice as long, as is the 22-year
period in the sun).
The magnetic field has no dynamical effects; it regulates the
mass loss rate by the formation of magnetic cool spots.
 The enhanced magnetic activity at the cycle maximum results in more
magnetic cool spots, which facilitate the formation of dust, hence
increasing the mass loss rate.
 The strong magnetic activity implies that the AGB star is spun up
by a companion, via a tidal or common envelope interaction.
 The strong interaction with a stellar companion explains the
observations that the concentric semi-periodic shells are found
mainly in bipolar PNs.

\noindent
{\it Subject heading:}   % to APJ
% {\it Key words:}         % to MNRAS
  Planetary nebulae: general
$-$ stars: AGB and post-AGB
$-$ stars: mass loss
$-$ stars: magnetic fields 
$-$ circumstellar matter

%\clearpage 

% ======================================================================
\section{INTRODUCTION}
% ======================================================================

 Concentric semi-periodic shells (also termed arcs or rings) appear in
the images of several asymptotic giant branch (AGB) stars,
proto-planetary nebulae (PNs) and young planetary nebulae (PNs).
 PNs and proto-PNs known to possess such shells are
CRL 2688 (the ``Egg'' nebula; Sahai {\it et al.} 1998);
IRAS 17150-3224 (Kwok, Su, \& Hrivnak 1998);
IRAS 17441-2411 (Su {\it et al.} 1998); 
Roberts 22 (although the shells are not really circular; Sahai {\it et al.}
1999); NGC 7027 and NGC 6543 (Bond 2000); and HB5. 
 Presently the only AGB star known to possess shells is
IRC+10216 (Mauron \& Huggins 1999).

The main properties of the shells are
(1) they are semi-periodic with time intervals between consecutive
ejection events of $\sim 200-1,000 \yrs$.
(2) They are spherical or almost spherical. However, low degree of
 departures from sphericity is seen in some shells, e.g., in the Egg
 nebula (Sahai {\it et al.} 1998).
(3) They can be almost complete, i.e., appear as rings,
or incomplete, where only a fraction of a full circle is observed,
i.e., they appear as arcs.
(4) The shells' density enhancement relative to the inter-shell density
is by a factor of a few up to a factor of
$\sim 10$ (Mauron \& Huggins 1999).
(5) The centers of all shells in a given object coincide with the
central star to within a few percent of their size.
(6) {\it All} PNs and proto-PNs which possess concentric shells
are bipolar, i.e., have two lobes with an equatorial waist between them
(e.g., IRAS 17150-3224), or they are extreme ellipticals (NGC 6543).
 By extreme elliptical PNs I refer to PNs having strong concentration of
mass toward the equatorial plane, e.g., a torus (ring).
 Although there might be a selection effect in detecting the shells in
bipolar PNs, since the central star is more attenuated (R. Sahai, private
communication), I do not think this alone can explain the observations.

 In the present paper I propose that these shells are produced by a
solar-like cycle in the progenitor AGB stars.
 The enhanced magnetic activity at the cycle maximum results in more
magnetic cool spots, which facilitate the formation of dust, hence
increasing the mass loss rate (Soker 2000).
 In $\S 2$ I review the previous mechanisms proposed for the formation
of these shells, and argue that none of these can account for all
properties of these shells.
I then outline the main ingredient of the magnetic activity cycle
mechanism.
 In $\S 3$ I examine some of the properties of a plausible
mechanism that may amplify the magnetic field in upper AGB stars.
 My summary is in $\S 4$.

% ======================================================================
\section{SUPPORT AND CONSTRAINTS FROM OBSERVATIONS}
% ======================================================================
% ========================================
\subsection{Previously Proposed Mechanisms}
% ========================================

 A discussion of several possible mechanisms for the formation of
concentric semi-periodic shells is given by Sahai {\it et al.} (1998)
and Bond (2000).
  Here I extend these discussions and examine each of the previously
proposed models.
\newline
{\bf Helium-shell flashes (thermal pulses).}
 Any mechanism based on helium-shell flashes is ruled out because
the typical inter-flash period is $\gtrsim 10^4 \yrs$
(Sahai {\it et al.} 1998; Kwok {\it et al.} 1998). 
\newline
{\bf Instability in the dust$+$gas outflow.} The instability in the
gas-dust coupling in the outflowing material was suggested as a
mechanism for the formation of the shells in the Egg nebula
by Deguchi (1997).
 This model cannot explain the formation of shells for a few reasons.
 First, from the results of Morris (1992) it seems that in
most cases the time interval between consecutive shells predicted by this
mechanism is too short.
Second, the shells will be smoothed out within a short distance
from the star (Mastrodemos, Morris \& Castor 1996).
 Third, the instability in the gas-dust coupling is  a local instability.
 Therefore, it will form small-scale instability and short arcs, but
will not form an almost complete shell (e.g., NGC 6543).
\newline
 {\bf Chaos.} Icke, Frank \& Heske (1992) examined the response
of the outer layers of an evolved AGB star to the oscillatory
motion of an instability zone in the stellar interior.
  They found that for the right initial conditions and parameters,
the stellar surface shows multiperiodicity or chaotic behaviors,
in addition to the regular oscillations.
 I find some problems with this mechanism.
Qualitatively, the chaotic or multiperiodicity found by Icke {\it et al.}
(1992) do not have the correct behavior (the two lower panels of
their fig. 11).
 There is no real semi-periodic behavior, but rather the time intervals
between two consecutive high amplitudes episodes differ a lot
from one interval to another.
 This behavior cannot account for the regularly spaced arcs in, e.g., 
IRAS 17150-3224 (Kwok {\it et al.} 1998).
 In some cases the duration of the maximum phase is longer than the
duration of the low amplitude intervals.
 This is not the observed properties of the concentric shells in most of
the objects listed in the previous section.
 In the lower two panels of their figure 11 (panels 7 and 8), the
maximum time interval between two consecutive maximum phases is only
$\sim 16$ times as long as the regular oscillation period.
 This time interval is an order of magnitude shorter than the observed
time intervals. 
 In panels 5 and 6 of their figure 11 the maximum phase lasts several
hundred years.
 However, they do not show more than one maximum phase, so
I can not comment on the long term behavior of these cases. 
\newline
{\bf  A binary companion in an eccentric orbit.}
In this mechanism, which was proposed by Harpaz, Rappaport \& Soker
(1997), a periastron passage of a stellar companion in an eccentric orbit
modulates the mass loss rate and/or geometry.
The periodic, on a time scale of several$\times 100 \yrs$,
periastron passage can increase, or decrease by diverting the flow,
the mass loss rate, leading to the formation of rings
by these periodic modulations.
 Sahai {\it et al.} (1998) criticized this mechanism on the ground that
it predicts exact circular shells with regular spacing between them,
properties which are not observed in the Egg nebula.
 There is another reason to reject this mechanism for the formation of the
shells.
 The eccentric orbit mechanism predicts that the center of the
shells will be displaced from the central star
(Soker, Rappaport, \& Harpaz 1998).
Such a displacement is not observed.
\newline
{\bf  A close binary companion.}
 Mastrodemos \& Morris (1999) show in their numerical simulations
that the presence of a close companion, orbital separation of
$\lesssim$several$\times 10 \AU$ leads to the formation
of a spiral structure in the equatorial plane.
 When viewed at a large angle to the symmetry (rotation) axis, the
circumstellar matter should show regularly spaced half-rings on
each side of the symmetry axis.
  I find this model unsatisfactory since it predicts on-off locations
for the half-shells near the symmetry axis.
  That is, the dense rings on one side of the symmetry axis will be
at radial distances which correspond to the inter-ring spaces on
the other side.
 This is not observed.
In addition, the arguments listed by Sahai {\it et al.} (1998)
against the eccentric binary mechanism hold for this mechanism as well.
\newline
{\bf  Giant convection cells.} Sahai {\it et al.} (1998) present the idea
that large cool convection cells form the concentric semi-periodic
shells.
 Such giant cool convection cells make dust formation
more efficient, hence increasing the mass loss rate.
I see two problems. First, giant convection cells are expected to
appear in specific location on the surface, so it is not clear they can
form an almost complete shell.
 Second, the time scale of several hundred years is much too long for
a life time of even a large convection cell in AGB stars. 

 To summarize, all the mechanisms listed above are expected to have some
signatures on some nebulae formed from AGB stars, but none of them
can explain the concentric semi-periodic shells.

% ================================================
\subsection{The Proposed Magnetic Cycle Mechanism}
% ================================================

 I conjecture that the mechanism behind the formation of the
concentric semi-periodic shells is a solar-like magnetic activity cycle.
 Below I list the observations in support of the proposed mechanism, the
basic processes of the mechanism, and the implications
of this conjecture.
In the next section I will elaborate on plausible
dynamo processes to amplify the magnetic field.

\subsubsection{Supporting Observations}

\noindent {\it 1) Magnetic fields in AGB stars.}
Kemball \& Diamond (1997) detected a magnetic field in the extended
atmosphere of the Mira variable TX Cam. 
 Kemball \& Diamond find the intensity of the magnetic 
field in the locations of SiO maser emission,
at a radius of $4.8 \AU \simeq 2 R_\ast$, to be $B \lae 5 G$.
The detection of X-ray emission from a few M giants (H\"unsch {\it et al.} 
1998) also hints at the presence of magnetic fields in giant stars. 
\newline
{\it 2) Solar cycle.} From the solar cycle we know that magnetic activity 
can be semi-periodic, and possess a global pattern, i.e., the cycle 
affects the entire solar surface. 
\newline
{\it 3) Inhomogeneity.} We also know from the solar magnetic activity that
the magnetic spots cover only a fraction of the solar surface. 
This inhomogeneity can explain incomplete shells and the inhomogeneity
observed in many shells, e.g., in the Egg nebula.
\newline
{\it 4) Spot distribution with latitude.}
 From the Maunder's butterfly diagram, e.g.,  for the years 1954-1977
(Priest 1987), there is evidence that the spots' distribution is most uniform
when the number of spots is at maximum, and the spots reach the
highest latitude during this cycle maximum.
 At that phase the spots are distributed almost uniformly from close to the
equator up to a latitude $\theta_m$.
 Spots do not distribute from the equator to $\theta_m$
in other phases of the solar cycle;
at the beginning of a cycle the are concentrated in two annular regions
around latitudes $\sim \pm 30 ^\circ < \theta_m$, while toward the end
of a cycle they are near the equator.
 Moreover, from the two solar cycles in these years it turns out that
$\theta_m$ is larger when the maximum total number of spots is larger.
 This hints that for  a very strong magnetic activity, i.e., 
when there are many spots, as I speculate is the case for the AGB 
progenitors of the concentric semi-periodic shells, the 
spots are distributed uniformly, up to the inhomogeneity 
discussed above, over the entire stellar surface.

\subsubsection{Basic Processes}
 The processes by which magnetic activity regulates the mass loss
rate from AGB stars are studied in earlier papers
(Soker 1998, 2000; Soker \& Clayton 1999).
 As in the sun, it is assumed that the magnetic activity leads to the
formation of magnetic cool spots, which facilitate the formation of dust.
 Since the mass loss mechanism from AGB stars is radiation pressure
on dust, higher magnetic activity leads to enhanced mass loss rate.
 The goal in the earlier papers was to explain the transition from
spherical to axisymmetrical mass loss in the AGB progenitors of
elliptical PNs.
 The idea is that the increase in the magnetic activity
(Soker \& Harpaz 2000) and/or the increase in the mass loss rate
(Soker 2000) which occur as the star is about to leave the AGB,
increase the mass loss rate in the equatorial plane more than they do
in the polar directions.
 This is based on the assumption, following the behavior of the sun,
that the dynamo magnetic activity results in the formation of more
magnetic cool spots near the equatorial plane than near the poles.
 Since that mechanism for axisymmetrical mass loss is intended to
explain the formation of elliptical PNs, it is also assumed
that the progenitors of elliptical PNs are slow rotators (Soker 2000),
having angular velocities in the range of
$3 \times 10^{-5} \omega _{\rm Kep} \lesssim \omega \lesssim 10^{-2}
\omega_{\rm Kep}$, where $\omega_{\rm Kep}$ is the equatorial
Keplerian angular velocity.
 Such angular velocities could be gained from a planet companion of
mass $\gae 0.1 M_{\rm Jupiter}$, which deposits its orbital angular
momentum to the envelope, or even from single stars 
which are fast rotators on the main sequence.   

 In the present case, the AGB stars are progenitors of bipolar
PNs or extreme ellipticals.
 In the binary model for the formation of bipolar PNs most of the
AGB progenitors are tidally spun-up by close companions
(Soker \& Rappaport 2000), while the extreme elliptical PNs may be
formed through a common envelope interaction (Soker 1997).
 In both cases we expect the AGB star to rotate with angular velocity of
\begin{equation}
0.01 \omega_{\rm Kep} \lesssim \omega \lesssim 0.1 \omega_{\rm Kep},
\end{equation}
which is more than the value of $\omega/\omega_{\rm Kep}$ in the sun.
 We expect strong magnetic activity since the convection motion is
very strong in these stars (next section).
 The upper limit on the angular velocity means that dynamical effects
will not much influence the mass loss geometry, since the centrifugal
forces are negligible.

\subsubsection{Implications}
 For the proposed mechanism to explain the spherical shape of the
shells, whether complete or not, the following are implied.
\newline
{\it 1) Spherical magnetic activity.} The average concentration of
magnetic cool spots should be uniform on the stellar surface, although
at any given moment the number of spots can be non-uniform, leading to
the small-scale nonuniformity of the concentric semi-periodic shells.
 We note that the mass loss rate during the formation of the inter shells
medium is not very high (the shells and the inter shells form a faint halo).
 It seems that large cool spots are required to regulate the mass loss
rate when the mass loss rate is low (Frank 1995; Soker 2000).
This means that only the medium to large cool magnetic spots are
required to be distributed uniformly, but not the small spots.
 Only when mass loss gets to be very high (for detail see Soker 2000),
as expected close to the termination of the AGB, do small cool spots
facilitate the formation of dust as well.
\newline
{\it 2) No other mechanisms.} The role of any other mechanism that causes
departure from spherical mass loss should be very small, e.g.,
nonradial pulsations.
 This implies that any detached binary companion cannot be too close,
i.e., no Roche lobe overflow, and that the AGB star cannot
rotate at $\omega \gtrsim 0.1 \omega_{\rm Kep}$.
 This is indeed the case in most progenitors of bipolar PNs, as discussed
in $\S 2.2.2$ above.
  The requirement that the companion spins up the mass losing star, but
have only a minor dynamical influence on the mass loss process, puts
severe constraints on the the binary properties.
 Mainly, the companion should not form an accretion disk and blow
a collimated fast wind (CFW) or jets when the shells are formed.
 This implies that the companion is likely to be a main sequence
star of mass $0.1 \lesssim M_2 \lesssim 0.5 M_\odot$, (for the conditions for
the formation of a CFW see Soker \& Rappaport 2000).
 Only during the superwind phase, when mass loss rate is very high,
does the companion manage to blow a CFW, leading to the
formation of a bipolar PN (Soker \& Rappaport 2000).
 The constraints on the companion mass, of $\sim 0.3 M_\odot$, and 
on the orbital separation, of $a \sim 5-30 AU$, explain why many 
bipolar PNs and proto-PNs do not have concentric semi-periodic shells.

% ======================================================================
\section{THE DYNAMO IN AGB STARS}
% ======================================================================

 Dynamo generation of magnetic fields in evolved AGB stars has
two major differences from the dynamo in main sequence stars, e.g.,
the sun.
 First the mass loss rate is very high, so that the mass leaving the star
drags the magnetic field lines, rather than being dragged by the
magnetic field lines.
 Second, the dynamo number is $N_D \ll 1$, whereas in main sequence
stars $N_D > 1$, as is required by standard $\alpha \omega$ dynamo models.
 The dynamo number is the square of the ratio of the magnetic field
amplification rate in the $\alpha \omega$ dynamo model, to the ohmic
decay rate. 
 A third difference from the situation in the sun, but not from
all main sequence stars, is that in AGB stars the convective region
is very thick, whereas in the sun its width is only $ 0.3 R_\odot$. 
 The aim of this section is to point to possible effects which
these differences may have on the amplification of the magnetic field,
and not to develop a new dynamo mechanism.
 Future more complete calculations should examine the exact conditions
and mechanism(s) for the generation of magnetic fields in AGB stars.

% ======================================================================
\subsection{Effects Due to a High Mass Loss Rate}
% ======================================================================

 In the sun, as in most main sequence stars, the mass loss rate
is determined mainly by the magnetic activity.
The magnetic pressure $P_B=B^2 /( 8 \pi)$ on the stellar surface is
no less than the ram pressure of the wind $P_w=\rho v_w^2$,
where $\rho= \dot M_w /(4 \pi R^2 v_w)$ is the density, $v_w$ is the
wind velocity, $R$ is the stellar radius, and $\dot M_w$ is the mass
loss rate to the wind, defined positively.
 Substituting typical values for the sun
we find $P_B \simeq 0.1 (B/2 \G)^2 \erg \cm^{-3}$ and
$P_w \simeq 10^{-3}  \erg \cm^{-3}$, hence 
$P_B/P_w \simeq 100$. 
 The relative magnetic activity required to dictate the mass loss 
geometry from AGB stars via the enhanced dust formation above
magnetic cool spots is much weaker, and can be as low as 
$P_B/P_w \simeq 10^{-4}$ (eq. 11 of Soker 1998).
  Therefore, while in main sequence stars the magnetic field drags the wind
close to the stellar surface, in AGB stars the wind drags the magnetic
field lines.
 Assuming that the wind conserves angular momentum, its angular velocity
decreases as $(R/r)^2$, where $r$ is the distance of
a parcel of gas from the center of the star.
 Hence near the surface
\begin{equation}
\left( \frac { d \omega}{d r } \right)_{\rm surface} 
= - \frac {2 \omega}{R}.
\end{equation}
 Note that in the solar interior  the differential rotation is weaker
$\vert d \omega / dr \vert \lesssim \omega_\odot / R_\odot$
(Tomczyk, Schou, \& Thompson 1995; Charbonneau {\it et al.} 1998).
 Even if on the surface of AGB stars the shear is lower than the shear
in the inner boundary of the convection region, it occurs on a much
larger area, since the inner boundary of AGB convective regions is at
several$\times R_\odot$.

 Although the angular velocity shear is similar at the equator and poles,
the winding of the field lines will be much stronger near the equator.
 Winding will be at large angles only when the wind is not much
faster than the rotation velocity near the equator.
 This is indeed the case for the AGB stars considered in this paper, 
for which the angular velocity is according to equation (1).
 For a $1 M_\odot$ AGB star with a stellar radius of $R=2 \AU$, 
equation (1) gives for the rotation velocity on the equator  
$0.2 \km \s^{-1} {\lesssim} {\it v}_{eq} \lesssim 2 \km \s^{-1}$.
 The distance along which the wind from upper AGB stars is accelerated is
$\sim R$, and therefore within this distance from the surface the wind
velocity is $< 10 \km \s^{-1}$, with a much lower velocity  
just above the surface: $ v_{ws} \ll 10 \km \s^{-1}$. 
 The magnetic field lines on the surface will be inclined in the azimuthal
direction at an angle $\alpha$ to the radial direction,
which depends on the latitude $\theta$
($\theta =0$ at the equator) according to 
\begin{equation}
\tan \alpha (\theta) =  (v_{eq}/v_{ws}) \cos \theta.  
\end{equation}
   
  The conclusion from this subsection is that the amplification of the 
magnetic field at the AGB stellar surface, i.e., the outer boundary of 
the convection region, may be more significant that at the inner boundary 
of the convection regions of AGB stars.

% ======================================================================
\subsection{Small Dynamo Number}
% ======================================================================

 In the $\alpha\omega$ stellar dynamo mechanism the $\alpha$
effect, due to convection, generates the poloidal component of
the magnetic field, while differential rotation generates the
toroidal component (e.g., Priest 1987 and references therein).
  Theory predicts that this mechanism operates efficiently only
when the dynamo number  (which is the square of the ratio of the 
magnetic field amplification rate to the ohmic decay rate of the 
magnetic field) is $N_D>1$.
 When comparing with observations it is convenient to use the
Rossby number (Noyes {\it et al.} 1984; Saar \& Brandenburg 1999).
 The Rossby number is proportional to the ratio of the rotational period,
$P_{\rm rot} = 2 \pi / \omega$, to the convective overturn
time $\tau_c$.
 Following  Noyes {\it et al.} (1984) I take $\tau_c= 2 l_p/v_c$,
where $l_p$ is the pressure scale height and $v_c$ is the convective
velocity, hence
${\rm Ro} \equiv (\omega \tau_c)^{-1}= P_{\rm rot} v_c/(4 \pi l_p) $.
 Noyes {\it et al.} (1984) based their use of the Rossby number
on the crude approximate relation $N_d \sim {\rm Ro}^{-2}$.
 For main sequence stars having magnetic activity
${\rm Ro} \lesssim 0.25$ (Saar \& Brandenburg 1999),
and hence $N_D > 1$ as required for the $\alpha\omega$ dynamo mechanism.
 For the sun ${\rm Ro}_\odot = 0.16$, while
the values of the Rossby number for the superactive stars in the
sample used by Saar \& Brandenburg (1999) are in the range
$5 \times 10^{-5} \lesssim {\rm Ro} \lesssim 10^{-2}$.
 Because of the strong convection in the envelope of AGB
stars we find that ${\rm Ro}(AGB) \gg 1$.
  Using typical values for AGB stars (e.g., figs 1-5 of
Soker \& Harpaz 2000; note that the density in their figs. 1-5 is lower by
a factor of 10; the correct density scale is in their fig. 6),
we find the Rossby number to be
\begin{equation}
{\rm Ro} (AGB) = 9
\left( \frac {v_c } {10 \km \s^{-1}} \right)
\left( \frac {l_p } {40 R_\odot} \right)^{-1}
\left( \frac {\omega} {0.1 \omega_{\rm Kep}} \right)^{-1}
\left( \frac {P_{\rm Kep} } {1 \yr } \right),  
\end{equation}
where $P_{\rm Kep}$ is the orbital period of a test particle moving
in a Keplerian orbit along the equator of the star.
 The low value of the dynamo number, $N_D \sim {\rm Ro}^{-2} \ll 1$, 
suggests that the convective motion amplifies both the poloidal and 
toroidal magnetic components, but that the differential rotation, 
both inside the envelope and on the surface (see previous subsection), 
still plays a nonnegligible role. 
  Hydrodynamic turbulence can amplify magnetic fields,
although not as efficiently as the $\alpha\omega$ dynamo
(see, e.g., Goldman \& Rephaeli 1991, and references therein, for 
the amplification of magnetic fields in clusters of galaxies). 
 Taking into account the observations that suggest the presence
of magnetic fields in AGB stars ($\S 2.2.1$ above), I conclude that
the strong convective motion in AGB stars together with the rotation
can indeed amplify the magnetic field via an $\alpha^2 \omega$ dynamo. 
 
 In previous papers I argued that this $\alpha^2 \omega$ dynamo can
operate, although at a low activity level, even in AGB stars rotating 
as slowly as $\omega \simeq 10^{-4} \omega_{\rm Kep}$ (Soker 1998), 
and in some cases even as low as 
$\omega \simeq 3 \times 10^{-5} \omega_{\rm Kep}$ (Soker \& Harpaz 2000).
  We note that with this angular velocity the 
mass loss time scale 
$\tau_m = M_{\rm env} / \vert \dot M_{\rm env} \vert$, where
$M_{\rm env}$ is the envelope mass, is not much shorter than $1/\omega$.
 Substituting typical values for AGB stars which are expected to be 
the progenitors of elliptical PNs, during their super-wind phase
\begin{equation}
\frac {\omega ^{-1}}{\tau_m} = 1.6 
\left( \frac {\omega }{10^{-4} \omega_{\rm Kep}} \right)^{-1}
\left( \frac {P_{\rm Kep}} { 1 \yr} \right)
\left( \frac {M_{\rm env}}{0.03 M_\odot} \right)^{-1}
\left( \frac {\vert \dot M_{\rm env} \vert}
{3 \times 10^{-5} M_\odot \yr^{-1}} \right).
\end{equation}
 This supports the assumption that the angular velocity plays a 
nonnegligible role in the $\alpha^2 \omega$ dynamo even in these 
very slowly rotating AGB stars. 
 However, the magnetic activity is expected to be weak, and the magnetic
cool spots to be concentrated in and near the equatorial plane 
(see $\S 2.2.1$ above). 
In these stars, contrary to the case with the stars discussed in the
present paper, the rotation is too slow to excite magnetic activity 
close to the poles, hence a higher mass loss rate near the equatorial 
plane leads later to the formation of an elliptical PN.

% ======================================================================
\section{SUMMARY}
% ======================================================================
 
  In this paper I propose that the concentric semi-periodic shells 
found around several PNs, proto-PNs, and AGB stars are formed by a
magnetic activity cycle in upper AGB stars. 
 The main assumptions, processes and implications of the
proposed mechanism for the formation of the shells are 
listed below, together with the explanations for the shells' properties
listed in section 1. 
\newline
{\bf (1)} It is assumed that the magnetic activity leads to the formation
of a large number of magnetic cool spots.
The cool spots enhance dust formation (Soker 2000 and references
therein), and hence increase the mass loss rate. 
The magnetic field, though, has no dynamical effects; its only
role is to form cool spots.
The formation of dust above cool spots is a highly nonlinear process
(Soker 2000), and therefore a relatively small increase in the number
of cool spots will substantially increase the mass loss rate.
 This explains the observations that the shells are much denser
than the inter-sells density,  by up to a factor of $\sim 10$. 
\newline
{\bf (2)} The sporadic nature of the appearance of magnetic cool spots
on the stellar surface, as in the sun, explains the incompleteness of 
some shells and other small-scale shell inhomogeneities. 
\newline
{\bf (3)} The spherical shells mean that the magnetic cool spots are 
distributed uniformly over the entire AGB stellar surface (up to
the inhomogeneities mentioned above).
This is indeed expected for strong magnetic activity ($\S 2.2.1$).
\newline
{\bf (4)} The strong magnetic activity implies that AGB stars which form
concentric shells are relatively fast rotators,
$0.01 \lesssim (\omega/\omega_{\rm Kep}) \lesssim 0.1$.
They are spun up by a stellar companion via tidal interaction, or
via a common envelope phase.
 Such tidal interactions are likely to form bipolar PNs
(Soker \& Rappaport 2000), whereas a common envelope interaction
is likely to form an extreme elliptical PNs.
This explains why the shells are found in bipolar or extreme elliptical
PNs and proto-PNs. 
 Two things should be noted here:
($i$) The AGB stars cannot rotate too fast since then the centrifugal
force will become nonnegligible and the shells will not be spherical. 
It is indeed expected that $\omega < 0.3 \omega_{\rm Kep}$
in most progenitors of bipolar PNs (Soker \& Rappaport 2000). 
 The companion cannot blow a collimated fast wind, or jets, during the
phase of the shell formation. This constraints the companion
to be a main sequence, rather than a white dwarf, and of relatively
low mass $M_2 \sim 0.1-0.5 M_\odot$.
($ii$) In very slowly rotating AGB stars the magnetic activity is very 
weak, the cycle period is extremely long, and the cool spots are 
expected to be concentrated near the equator (see $\S 2.2.1$), 
hence leading to the formation of elliptical PNs 
(Soker 1998, 2000). 
\newline
{\bf (5)} It is also assumed that, as in the sun and other
main sequence stars, the magnetic activity has a semi-periodic
variation.
This is the explanation for the semi-periodic nature of the shells.
In main sequence stars the ratio of the period of the magnetic
activity cycle $P_{\rm cyc}$ to the rotation period is
(Baliunas {\it et al.} 1996; Saar \& Brandenburg 1999)
$ 50 \lesssim  P_{\rm cyc}/P_{\rm rot} \lesssim 10^5$.
 Therefore, for an AGB stellar rotation period of $\sim 10-100 \yrs$,
magnetic activity cycles of periods of $200-10^3 \yrs$ require this
ratio to be somewhat smaller  
$P_{\rm cyc}/P_{\rm rot} \sim 10$.
\newline
{\bf (6)} There is no dynamo model for AGB stars. 
 Based on some observations listed in $\S 2.2.1$, I assume that a dynamo
can indeed amplify magnetic fields in AGB stars.
 In the present paper I did not develop or calculate any dynamo mechanism 
for AGB stars. I only point here ($\S 3$) to the two major differences
between the standard $\alpha\omega$ dynamo mechanism for
main sequence stars and any dynamo mechanism for AGB stars. 
First, the high mass loss rate means that the wind drags the 
magnetic field lines in AGB stars, contrary to the case with main sequence
stars. 
This suggests that the azimuthal shear near the surface plays a role 
in the dynamo mechanism, as well as the shear in the stellar interior. 
Second, the dynamo number is $N_D \ll 1$ (or the Rossby number is
${\rm Ro} \gg 1$) in AGB stars, whereas the standard $\alpha\omega$ dynamo
mechanism requires $N_D >1$, as is the case with active main sequence stars.
 This suggests that the main amplification of magnetic fields in AGB 
stars is via the convective motion, but with a nonnegligible role
of the rotation, i.e., an $\alpha^2 \omega$ dynamo.
\newline
{\bf (7)} This mechanism proposed in the present paper
has some predictions.
$(i)$ It predicts that the AGB progenitors of the
concentric semi-periodic shells have main sequence
companions of mass $\sim 0.1-0.5 M_\odot$, with orbital 
periods in the rang of $\sim 15-150 \yrs$. 
 The orbital periods predicted by the binary models mentioned in $\S 2.1$,
on the other hand, are in the range of $\sim 200-10^3 \yrs$. 
$(ii)$ The proposed mechanism predicts that almost all PNs and proto-PNs 
with concentric semi-periodic shells are bipolar or extreme elliptical PNs. 
$(iii)$ In some case the shells can be formed after the mass
losing star was spun up via a common envelope evolution. 
 In these cases most of the descendant PNs are expected to be 
extreme elliptical PNs, rather than bipolar PNs, and  
either the companion has a final orbital period of less than a year, 
even only a few hours, or else the companion is
completely destructed in the common envelope.

{\bf ACKNOWLEDGMENTS:} 
 I thank Raghvendra Sahai for very helpful comments. 
 This research was supported in part by grants from the Israel Science
Foundations and the US-Israel Binational Science Foundation.

%\ newpage

\end{document}